%% file: main.tex
\documentclass[10pt, conference]{IEEEtran}
\IEEEoverridecommandlockouts
\usepackage{amsmath,amssymb,amsfonts}
\usepackage{algorithmic}
\usepackage{graphicx}
\usepackage{tabularx}
\usepackage{textcomp}
\usepackage{xcolor}
\usepackage{siunitx}
\usepackage{booktabs}
\usepackage{url}
\usepackage{enumitem}
\usepackage{diagbox}
\usepackage{multicol}
\usepackage{multirow}
\usepackage{threeparttable}
\usepackage[numbers]{natbib}
\usepackage[listings]{tcolorbox}
\usepackage[misc, geometry]{ifsym}

\usepackage{pifont}% http://ctan.org/pkg/pifont

\input{macro} %Jianjun Zhao (20210106)
\usepackage{braket}
\usepackage {colortbl} 
\def\BibTeX{{\rm B\kern-.05em{\sc i\kern-.025em b}\kern-.08em
    T\kern-.1667em\lower.7ex\hbox{E}\kern-.125emX}}
\begin{document}

\title{QChecker: Detecting Bugs in Quantum Programs via Static Analysis\\
%\title{QChecker: A Static Analysis Tool for Quantum Programs\\
}

\author{
\IEEEauthorblockN{Pengzhan Zhao,
Xiongfei Wu,
Zhuo Li,
Jianjun Zhao\IEEEauthorrefmark{1}
\thanks{\IEEEauthorrefmark{1}Jianjun Zhao (zhao@ait.kyushu-u.ac.jp) is the corresponding author.} 
}

\IEEEauthorblockA{
Graduate School and Faculty of Information Science and Electrical Engineering\\
Kyushu University, Japan\\
%Email: {zhao@ait.kyushu-u.ac.jp}
%{\{zhao.pengzhan.813, xiongfei.wu.a94, li.zhuo.786\}@s.kyushu-u.ac.jp, zhao@ait.kyushu-u.ac.jp}
%{zhao.pengzhan.813, xiongfei.wu.a94, li.zhuo.786\}@s.kyushu-u.ac.jp, zhao@ait.kyushu-u.ac.jp}
}
}

\maketitle

\begin{abstract}
% Static analysis supports checking code without running code, which has become increasingly mature in classical software testing. However, considering quantum properties, traditional tools are unsuitable for detecting quantum programs.
% %and there is no static analysis tool for debugging quantum programs. 
% In this paper, we propose QChecker, a python-based static analysis tool for detecting quantum programs.
% %It is the first static analysis tool oriented to detect quantum programs.
% QChecker contains two parts: program information extraction and bug-pattern-based detectors.
% We use the Bugs4Q benchmark suite to evaluate the performance of QChecker. 
% From the results, we found that QChecker can be well used to detect Qiskit programs.
% It also supports detecting programs in other common python-based quantum programming languages.
Static analysis is the process of analyzing software code without executing the software. It can help find bugs and potential problems in software that may only appear at runtime. Although many static analysis tools have been developed for classical software, due to the nature of quantum programs, these existing tools are unsuitable for analyzing quantum programs. This paper presents QChecker, a static analysis tool that supports finding bugs in quantum programs in Qiskit. QChecker consists of two main modules: a module for extracting program information based on abstract syntax tree (AST), and a module for detecting bugs based on patterns. We evaluate the performance of QChecker using the Bugs4Q benchmark. The evaluation results show that QChecker can effectively detect various bugs in quantum programs. 
%Please refer to the code base\footnote{\url{https://github.com/Z-928/QChecker}} for implementation details.

%The result shows that QChecker can basically find bugs in the Qiskit program correctly.
%QChecker is publicly available at:{\color{red}?}.
\end{abstract}


\iffalse
\input{sections/intro}
\input{sections/back.tex}
\input{sections/method.tex}
\input{sections/experiment.tex}
\input{sections/threats.tex}
\input{sections/related-work.tex}
\input{sections/conclusion.tex}
\fi

\section{Introduction}
%Recent advances in hardware have inspired the innovation of quantum software engineering~\cite{zhao2020quantum,piattini2020quantum} from both industry and academia. 
Quantum computing has been applied to many cutting-edge areas, such as quantum machine learning~\cite{biamonte2017quantum, dunjko2016quantum}, big data analysis~\cite{rebentrost2014quantum}, and molecular simulations~\cite{grimsley2019adaptive} due to its unique promising advantage over classical computing. 
Quantum programming is designing and constructing executable quantum programs to achieve a specific computational result.
With the rapidly growing complexity of quantum programs, it is decisive to alleviate the efforts in programming such quantum programs. 
%\textcolor{blue}{Quantum computing frameworks are for this rapidly emerging field of what deep learning (DL) frameworks and execution environments are for DL programs.} 
Several quantum programming languages are available for quantum programmers, such as Qiskit~\cite{aleksandrowicz2019qiskit}, Sliq~\cite{bichsel2020silq}, Cirq~\cite{cirq2018google}, Quipper~\cite{green2013quipper}, and Q\#~\cite{svore2018q}, allowing researchers and developers to implement and experiment with various quantum computing techniques quickly. 

Given the importance and wide application of quantum programming, ensuring the correctness of quantum programs is crucial for quantum software development. However, recent empirical studies~\cite{campos2021qbugs,paltenghi2021bugs,zhao2021identifying,zhao2021bugs4q} have shown that the current quantum program development process is still error-prone. While debugging and testing quantum programs has gained significant attention~\cite{ali2021assessing,fortunato2022mutation2,honarvar2020property,huang2019statistical,li2020projection,long2022testing}, the existing debugging and testing techniques often require dynamic execution of the underlying quantum programs. Considering that most of the current quantum programs are executed on quantum computers and simulators available on the cloud, the existing debugging and testing techniques can be cumbersome and expensive.

%Static analysis tools are programs that aim at inferring program properties of other programs by examining the source of other programs without running them. 
%\textcolor{red}{Static analysis for classical programs has been extensively studied and widely adopted in modern development practices~\cite{radhika2014static, wang2022icpc}. Due to the advantage in speed and cost, static bug finders (also known as static analyzers) that can infer program properties without program execution have been extensively studied and widely adopted in modern development practices to find bugs in classical programs~\cite{flanagan2002extended,hovemeyer2004finding,rutar2004comparison,shen2008xfindbugs}.} 

In classical software development practice, static analysis techniques have been widely used to detect various types of bugs in classical programs due to their advantages in speed and cost ~\cite{flanagan2002extended,hovemeyer2004finding,rutar2004comparison,shen2008xfindbugs,wang2022icpc}. However, detecting bugs in quantum programs via static analysis can be challenging. Since quantum computation logic is expressed in quantum circuits, and the states of quantum registers are measured probabilistically, static analysis tools designed for classical programs struggle to detect mistakes in quantum programs. 

To bridge the gap, we present QChecker, a static analysis tool designed for detecting bugs in quantum programs, especially for Qiskit. The approach addresses the challenge above by first distilling a set of common bug patterns summarized from real quantum bugs in previous studies~\cite{zhao2021identifying,zhao2021bugs4q} and then constructing eight detectors for detecting these bug patterns in quantum programs. The whole process is non-trivial since the distilled bug patterns must carefully consider the domain-specific constraints of quantum computing to be accurate and effective.

We evaluate QChecker on Bugs4Q~\cite{zhao2021bugs4q}, a realistic benchmark consisting of 42 real-world buggy quantum programs. 
%\textcolor{red}{Furthermore, we evaluate QChecker on seven buggy quantum programs written in Cirq.} 
Experimental results show that QChecker can efficiently detect bugs in quantum programs.
%and \textcolor{red}{exhibit extendibility to common Python-based quantum programming languages. }
Furthermore, we discuss the extendability of QChercker for other Python-based quantum programming languages.

In summary, this work makes the following contributions:
\begin{itemize}[leftmargin=2em]
\setlength{\itemsep}{3pt}
\item We present the first bug detection tool dedicated to quantum programs in Qiskit. Using static analysis techniques, QChecker can generate diagnostic messages that assist developers in pinpointing potential bugs in their programs quickly.
\item We implement QChecker and evaluate its effectiveness and performance in a real-world Bugs4Q benchmark. The results show that QChecker can effectively detect various types of bugs in quantum programs. 
  % \item We distill a set of common bug patterns for quantum programs, which is the foundation of performing static analysis on quantum programs.
%  \item We propose a novel approach of decomposition that represents quantum programs using \CodeIn{QP\_Operation} and \CodeIn{QP\_Attribute}, which is the foundation of performing static analysis on quantum programs.
%  \item We construct eight detectors based on our proposed bug patterns to enable the first static analysis tool for quantum programs. 
%  \item We integrate QChecker with the widely used Qiskit framework and have tested its extendibility to other Python-based quantum programming languages.
  
  % \item It avoids the collapse of quantum states caused by prying into the internal quantum state.
  
  % \item It saves the time and resources needed to test quantum programs dynamically.
  
  % \item It has proven effective in finding detects and vulnerabilities in Qiskit programs.
\end{itemize}

The rest of the paper is organized as follows. Section~\ref{sec:background} provides some basics of quantum programming. 
Section~\ref{sec:methodology} describes our QChecker approach for static analysis of quantum programs. Section~\ref{sec:experiments} presents the performance of QChecker on Bugs4Q. Section~\ref{sec:threats} reviews our threats of validity.
Section~\ref{sec:related} discusses related work, and Section~\ref{sec:conclusion} finally concludes this paper.

\section{Background}\label{sec:background}

We briefly introduce the background information on programming in Qiskit and the basic concepts of qubits.

\subsection{Qiskit}

Qiskit~\cite{aleksandrowicz2019qiskit} is one of the most widely used open-source frameworks for quantum computing, allowing us to create algorithms for quantum computers. As a Python package, it provides tools for creating and manipulating quantum programs and running on prototype quantum devices and simulators and can use built-in modules for noise characterization and circuit optimization to reduce the impact of noise. Qiskit also provides a library of quantum algorithms for machine learning, optimization, and chemistry.
%Qiskit provides a complete set of tools needed to interact with quantum systems and simulators and 

In Qiskit, a program is defined by a quantum object data structure that contains configuration information and the experiment sequences. The object can be used to get status information and retrieve results~\cite{mckay2018qiskit}.
Figure~\ref{fig:mesh1} shows a simple Qiskit program that illustrates the entire workflow of a quantum program. 
%In Qiskit, we need first to create a backend simulator \CodeIn{qasm\_simulator} as follows.
The function \CodeIn{Aer.get\_backend('qasm\_simulator')} returns a backend object for the given backend name (\CodeIn{qasm\_simulator}). 
The \CodeIn{backend} class is an interface to the simulator, and the actual name  of \CodeIn{Aer} for this class is \CodeIn{AerProvider}.
\noindent
After the experimental design is completed, the instructions are run through the \CodeIn{execute} method.
The \CodeIn{shots} of the simulation, which means the number of times the circuit is run, is set to 1000 while the default is 1024.
When outputting the results of a measurement, the method \CodeIn{job.result()} is used to retrieve the measurement results. We can access the counts via the method \CodeIn{get\_counts(circuit)}, which gives the experiment's aggregate outcomes.

%IBM offers an online platform, "IBM Quantum Experience~\cite{IBM}," which features two 5-qubit processors and one 16-qubit processor, making it very powerful. We can use Qiskit to build the circuit directly in the online backend. It is also possible to program locally using Qiskit in a standalone Jupyter programming environment. Another powerful aspect of Qiskit is that there is no need to do the manual work of unassigning qubits, making programming with Qiskit becomes much easier and more efficient.

\subsection{Basic Properties of Qubits}
In this subsection, we use Qiskit as an example to explain the characteristics of quantum bit (\textit{qubit} for short) and the necessary execution process of a complete quantum program.

\begin{figure}[t]
  \begin{CodeOut}
\footnotesize{
  \begin{alltt}
    simulator = \textbf{Aer.get_backend}("qasm_simulator")

    qreg = \textbf{QuantumRegister}(3)
    creg = \textbf{ClassicalRegister}(3)
    circuit = \textbf{QuantumCircuit}(qreg, creg)

    circuit.\textbf{h}(0)
    circuit.\textbf{h}(2)
    circuit.\textbf{cx}(0, 1)
    circuit.\textbf{measure}([0,1,2], [0,1,2])
    job = \textbf{execute}(circuit, simulator, shots=1000)
    result = job.\textbf{result}()
    counts = result.\textbf{get_counts}(circuit)
    \textbf{print}(counts)
  \end{alltt}
}
\end{CodeOut}
\Caption{\label{fig:mesh1}A simple quantum program in Qiskit}
\end{figure}

The basic unit of information in quantum computing is the qubit. As shown in Figure~\ref{fig:mesh1}, \CodeIn{qreg = QuantumRegister(3)} means assigning a quantum register of three qubits, and the value of each qubit is $\ket{0}$ by default. So the initial value of these three qubits is $\ket{000}$. Next, let the first and third qubits pass through the H (Hadamard) gate, as shown by \CodeIn{circuit.h(0)} and \CodeIn{circuit.h(2)}. In this way, the unique property {\it superposition} of qubits is realized, which means the qubit contains the states of $\ket{0}$ and $\ket{1}$.
There is also an {\it entanglement} of qubit properties that only multiple qubits can achieve. The code in the sample program is \CodeIn{circuit.cx(0,1)}. That is to say, the first qubit is entangled with the second qubit through a CNOT (Controlled-NOT) gate operation. We measure the first qubit, and its output is \CodeIn{0} for 50 percent probability and \CodeIn{1} for 50 percent probability. After that, measuring the second qubit is 100 percent the same as the first measurement result.
Since the third qubit is not related to the first two qubits, the last qubit's measurement result is still taken with \CodeIn{0} for 50 percent probability and \CodeIn{1} for 50 percent probability.
The measurement statement of qubits shown in Figure~\ref{fig:mesh1} is \CodeIn{circuit.measure([0,1,2], [0,1,2])}. Measurement can lead to the collapse of a quantum superposition state to a classical state. There are many kinds of quantum measurements, and the projection measurement of a single qubit is used here. That is, each qubit is projected onto a state space consisting of base vectors $\ket{0}$ or $\ket{1}$. In this program, the final output is a three-bit array.

\begin{figure*}[h]
    \centering
    \includegraphics[width=0.95\textwidth]{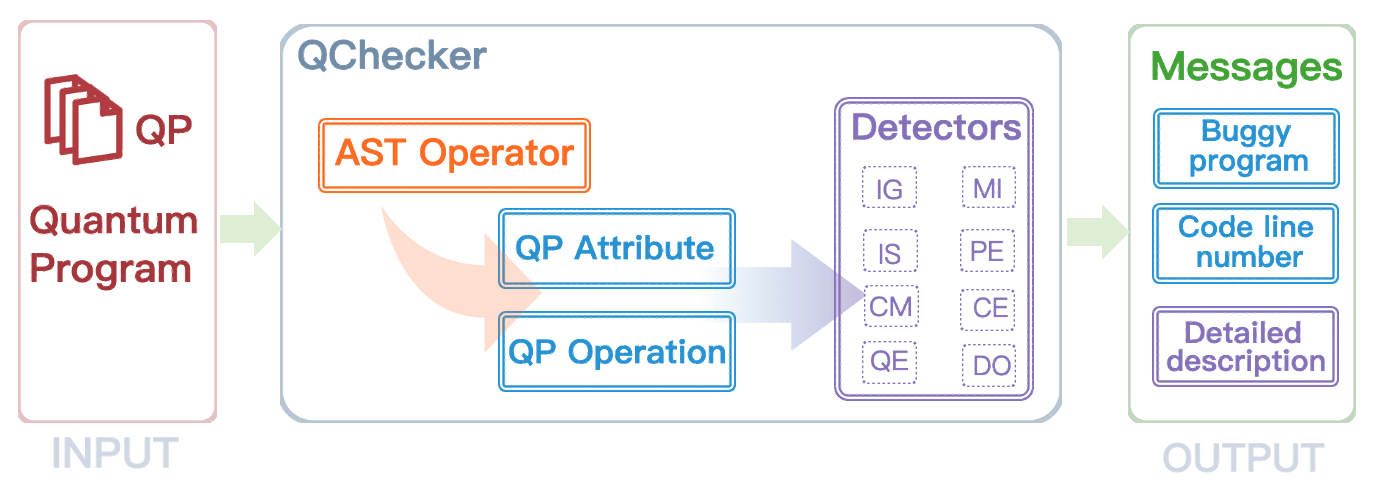}
    \caption{The structure of QChecker.}
    \label{fig:sketch}
\end{figure*}

\section{The QChecker Tool} \label{sec:methodology}
% In this section, we briefly introduce the construction of QChecker. 
% \Zhuo{In this section, we introduce the construction of QChecker. }
% QChecker is developed based on the Python language. We use the \textit{ast} module provided by Python to extract information from the program. The \textit{sys} and \textit{os} modules are used to implement QCheckcer's interaction with the system environment and the quantum program files under test. As well as the basic \textit{re} module to implement regular expression matching operations.
% Clearly, QChecker can achieve detection without the environment for executing quantum programs.

% %The QChecker tool is open-sourced at:{\color{red}?}.
% Figure~\ref{fig:sketch} shows how the QChecker finds bugs.
% First, the information of a quantum program is extracted by the \CodeIn{Ast\_operator} and decomposed into two parts: program attributes and program operation. Next, extracted information will be passed through the detectors.
% Finally, if bugs are found, QChecker will catch the bugs and provide information about the program and the error codes provided by detectors. If there is no bug detected, QChecker will return false.

In this section, we introduce the construction of QChecker, which is developed based on Python. As illustrated in Figure~\ref{fig:sketch}, QChecker first performs a thorough information extraction of the quantum programs based on their ASTs. The corresponding operations are in the module \CodeIn{Ast\_Operator}. The information mainly includes the variable assign operations and function calls, which will be further stored in \CodeIn{QP\_Attribute} and \CodeIn{QP\_Operation}. Then QChecker transmits the extracted information to the bug detectors. The bug detectors can detect various bug patterns, as shown in Table~\ref{table:1}.
Finally, QChecker generates bug detection reports, including the buggy programs, line numbers, and bug descriptions.

\subsection{Information Extraction}
% When QChecker receives a quantum program under test, the first thing is converting the program into the form of an abstract syntax tree (AST). 
% Next, QChecker provides two functions for information extraction: \CodeIn{QP\_Attribute} and \CodeIn{QP\_operation}. 

The previous static analysis tools inspire us (e.g., PyLint~\cite{thenault2001pylint}) that using AST for program information extraction is effective and efficient. However, different from classical static analysis tools, the \CodeIn{AST\_Operator} in QChecker has the ability to extract information specific to the semantics and the function of quantum programs. Taking the program shown in Figure~\ref{fig:mesh1} as an example, we apply a structured parsing to each quantum program file, i.e., generating the AST. We adopt two modules named \CodeIn{QP\_Attribute} and \CodeIn{QP\_Operation} to store the AST information of all the variables and function calls, respectively. In addition, QChecker also supports handling complex syntax and data structures such as dictionaries, lists, function definitions, loops, and conditional branches. The purpose of this design is that the structured AST-based information extraction can help QChecker trace the relationship between each variable and function call. For example, a variable may be modified multiple times, or its name may be changed when passed as an argument inside a function.
Nevertheless, we can still trace back the initial value of the variables in the program. We plot instances of \CodeIn{QP\_Attribute} and \CodeIn{QP\_Operation} in Figures~\ref{fig:attr} and \ref{fig:opt}, respectively. %The bug detectors can detect various bug patterns (e.g., IG, IS).

\begin{itemize}[leftmargin=2em]
\setlength{\itemsep}{3pt}
  \item \textbf{QP\_Attribute}: The AST structure of a variable includes its variable name, variable value (which can come from a constant, another variable, or the result of a function calculation), type, and location (line of code). As shown in Figure~\ref{fig:attr}, The \CodeIn{QP\_Attribute} module is designed in a key-value manner. The keys are variable names which can be the indices for the variable values. 
  \item \textbf{QP\_Operation}: The AST structure of a function call contains a list of its arguments, the type and value of each argument, its position, and other information. As shown in Figure~\ref{fig:opt}, The \CodeIn{QP\_Operation} module is a list that contains all the function calls in the quantum program file. In detail, each function call can be further divided into function call names, arguments, and values. This information is stored in a more comprehensive table from QChecker, which uses the function call strings in \CodeIn{QP\_Operation} as the index.
\end{itemize}

These two modules contain all the information of a quantum program and make it more straightforward for further bug detection.
Moreover, users can directly obtain the above information through QChecker based on the API we released. It is worth mentioning that those programs containing basic syntax errors (e.g., python indentation errors, unrecognized operators, undefined variables and functions, etc.) will not be processed by QChecker. Instead, they will be prompted as syntax errors and thus be excluded from the static checking.

\begin{figure}[t]
  \begin{CodeOut}
\footnotesize{
  \begin{alltt}
    ('simulator', '\textbf{Aer.get_backend}("qasm_simulator")') 
    ('qreg', '\textbf{QuantumRegister}(3)') 
    ('creg', '\textbf{ClassicalRegister}(3)') 
    ('circuit', '\textbf{QuantumCircuit}(qreg,creg)') 
    ('job', '\textbf{execute}(circuit,simulator,shots=1000)') 
    ('result', 'job.\textbf{result}()') 
    ('counts', 'result.\textbf{get_counts}(circuit)')
  \end{alltt}
}
\end{CodeOut}
\Caption{\label{fig:attr}Program information extracted by \CodeIn{QP\_Attribute}.}
\end{figure}

\begin{figure}[t]
  \begin{CodeOut}
\footnotesize{
  \begin{alltt}
    '\textbf{Aer.get_backend}("qasm_simulator")'
    '\textbf{QuantumRegister}(3)'
    '\textbf{ClassicalRegister}(3)' 
    '\textbf{QuantumCircuit}(qreg,creg)'
    'circuit.\textbf{h}(0)'
    'circuit.\textbf{h}(2)'
    'circuit.\textbf{cx}(0,1)'
    'circuit.\textbf{measure}([0,1,2],[0,1,2])'
    '\textbf{execute}(circuit,simulator,shots=1000)'
    'job.\textbf{result}()'
    'result.\textbf{get_counts}(circuit)'
    '\textbf{print}(counts)'
  \end{alltt}
}
\end{CodeOut}
\Caption{\label{fig:opt}Program information extracted by \CodeIn{QP\_Operation}.}
\end{figure}

\subsection{Bug Pattern Detectors}

Bug patterns are erroneous code idioms or bad coding practices that have been proven to fail time and time again, which are usually caused by the misunderstanding of a programming language’s features, the use of erroneous design patterns, or simple mistakes sharing common behaviors. Previous work has identified some bug patterns for the Qiskit programming language~\cite{zhao2021identifying,zhao2021bugs4q}. In this work, we refined these bug patterns and built eight detectors to detect them. Table~\ref{table:1} shows the name of detectors and descriptions of bug patterns. We briefly describe each detector as follows.
%and the bug patterns can be found.
%Detailed descriptions and functions of each detector are explained as follows:
\begin{table}[h]
%\begin{CodeOutFN}
\caption{\label{table:1} Bug patterns that each detector is responsible for.}
\footnotesize{
\begin{center}
\renewcommand\arraystretch{1.5}
\begin{tabular} {c|l}
\hline 

\rowcolor{lightgray}\textbf{Detector Name}&\textbf{Bug patterns Descriptions}\\

\hline
\multirow{3}{0em}{\textbf{IG}} 
& - Gates are not among the backend’s basis gates. \\
& - Handle custom multi-qubit gates.\\
& - Random gate is not defined. \\
%&Wrong uses of gates lead to incorrect reset \\

\hline
\multirow{1}{0em}{\textbf{MI}} 
& - Ignoring the effects of measurement. \\

\hline
\multirow{3}{0em}{\textbf{IS}} 
& - Number of qubits larger than the registers defined. \\
& - The insufficient number of qubits.\\
& - Insufficient length of classical registers. \\

\hline
\multirow{6}{0em}{\textbf{PE}} 
& - Instruction not in basis gates. \\
& - Incorrect parameters in gates.\\
& - Using classical bits for entanglement. \\
& - Same physical qubit used in one operation. \\
%& Parameter error during decontrol operation. \\
& - Not giving lists for coupling\_map. \\

\hline 
\multirow{4}{0em}{\textbf{CM}} 
& - Unrecognized parameters. \\
& - Quantum circuit interaction error.\\
& - Create redundant classical registers. \\
& - The wrong command was used. \\

\hline 
\multirow{4}{0em}{\textbf{CE}} 
& - Object call error. \\
& - Import error.\\
& - Backend error. \\
& - Translating error. \\

\hline 
\multirow{1}{0em}{\textbf{QE}} 
& - Issue with new from\_qasm\_str() method. \\
\hline

\multirow{1}{0em}{\textbf{DO}} 
& - Method has been deprecated. \\

\hline

\end{tabular}
\end{center}
}
%\end{CodeOutFN}

\end{table}

\subsubsection{Incorrect uses of quantum gates (IG)}
This detector mainly checks if quantum gates are called correctly.
It determines whether a gate is recognized by Qiskit and whether it has been defined. In addition, the compliance of a custom gate and a three-qubit gate with the specification would also be checked.

\subsubsection{Measurement related issue (MI)}
Incorrect measurement means not only the improper use of measure operation that cause bugs but also the wrong operation after measurement. As shown in Figure~\ref{fig:5}, the user wants to achieve a quantum teleportation program.
The measured qubits are used as control qubits to entangle with other qubits.
\begin{figure}[t]
  \begin{CodeOut}
\footnotesize{
  \begin{alltt}
    qc = \textbf{QuantumCircuit}(3, 3)
    qc.\textbf{x}(0)
    qc.\textbf{h}(1)
    qc.\textbf{cx}(1, 2)
    qc.\textbf{cx}(0, 1)
    qc.\textbf{h}(0)
    qc.\textbf{measure}(0, 0)
    qc.\textbf{measure}(1, 1)
    qc.\textbf{cx}(1, 2)       \textit{<- Problematic operation} 
    qc.\textbf{cz}(0, 2)       \textit{<- Problematic operation}
  \end{alltt}
}
\end{CodeOut}
\Caption{\label{fig:5}Example of Incorrect Mearsurement}
\end{figure}
This detector mainly acts after the \CodeIn{measure} method is called.
It iterates through the operations following the \CodeIn{measure} statement and determines whether the measured qubit appears as a control qubit in the double-qubit gate operations.

\subsubsection{Incorrect initial state (IS)} 
This detector does not simply check whether the definitions of \CodeIn{QuantumRegister} and \CodeIn{ClassicalRegister} conform to the specification. It determines whether the initialization satisfies the entire quantum program's operation on qubits.
Sometimes, the Qiskit program limits the number of qubits used when simulating quantum programs. Such as, after our validation, using \CodeIn{Aer.get\_backend('qasm\_simulator')} as backends supports less than 30 qubits for measure operation, while \CodeIn{BasicAer.get\_backend('qasm\_simulator')} supports less than 24 qubits.
In this case, the detector first checks the backend chosen by users and Identifies if the initialized qubits are out of limits.
When the number of initialized qubits is set to \CodeIn{n}, the checker will keep track of the number of qubits called in the program.

\subsubsection{Parameter error (PE)}
After a quantum gate is invoked, this detector is responsible for determining whether the parameters in the gate are correct, including the parameters that do not exist in multiple-qubit gates, and the wrong use of numeric types.
However, some bug patterns are not easy to find. From Figure~\ref{fig:6}, we can see that the user wants to assign the qubits in the register to the physical qubits, both \CodeIn{qreg[0]} are \CodeIn{qreg[5]} assigned to the physical qubits \CodeIn{12}. Therefore, the detector goes through the parameter values and checks for duplicate physical qubit occupancy.

\begin{figure}[t]
  \begin{CodeOut}
\footnotesize{
  \begin{alltt}      
qreg = qk.\textbf{QuantumRegister}(7)

\textbf{layout} = \{qreg[0]: 12,     \textit{<- Problematic operation}
          qreg[1]: 11,
          qreg[2]: 13, 
          qreg[3]: 17, 
          qreg[4]: 14, 
          qreg[5]: 12,     \textit{<- Problematic operation}
          qreg[6]: 6\}
  \end{alltt}
}
\end{CodeOut}
\Caption{\label{fig:6}Example of parameters error}
\end{figure}

\subsubsection{Command misuse (CM)} 
This detector could detect the wrong or improper use of commands. Sometimes, parameters are not recognized because the method name is miswritten.
On the other hand, some methods can not recognize parameters and raise errors. As shown in Figure~\ref{fig:7}, attribute \CodeIn{pulse.shiftphase()} is not in module \CodeIn{qiskit.phase}.
Some of the commands in Qiskit are difficult to detect. For example, there are more than two quantum circuits while the user wants to nest one circuit with the others: 1) Command \CodeIn{to\_gate()} could be used to change the circuit into a combination of gates embedded in other circuits. 
2) Command \CodeIn{decompose()} could be used to decompose the circuit for embedded operation automatically.

%can not be recognized by the circuit library.
%If the detector finds the unrecognized parameter, it will traverse the entire program to find the \CodeIn{decompose()}, which is needed by the circuit library.

\begin{figure}[t]
  \begin{CodeOut}
\footnotesize{
  \begin{alltt}      
phase = \textbf{Parameter}('phase')

\textbf{with} pulse.build(FakeAlmaden())\textbf{as} phase_test_sched:

        pulse.\textbf{shiftphase}(            \textit{<- Unrecognized}
        phase, pulse.\textbf{drive_channel}(0))

phase_test_sched.\textbf{instructions}
  \end{alltt}
}
\end{CodeOut}
\Caption{\label{fig:7}Example of command misuse.}
\end{figure}

\iffalse
\begin{figure}[t]
  \begin{CodeOut}
\footnotesize{
  \begin{alltt}      
ansatz = \textbf{RealAmplitudes}(num_qubits=1, reps=1)

for method in ['param_shift',
                'fin_diff', 
                'lin_comb']:
                
    grad = \textbf{Gradient}(method).       \textbf{<-} Unrecognized
           convert(CircuitStateFn(ansatz))
           
    \textbf{print}(f"{method} is ok")
  \end{alltt}
}
\end{CodeOut}
\Caption{\label{fig:7}Example of command misuse.}
\end{figure}
\fi

\subsubsection{Call error (CE)} 
This detector is responsible for call errors, including Python package calls, backend simulator calls, and translator calls.
Besides, the detector can check for problems with parameter declarations. As shown in Figure~\ref{fig:8}, this error is not a duplicate call to \CodeIn{PauliMeasurementBasis()}. After running the code, we found that there was an error of \textit{invalid qubits for basis}. The call of \CodeIn{PauliMeasurementBasis()} is invalid for \CodeIn{preparation\_basis}.
This types of bugs are hard to detect by QChecker, the detector can only judge one scenario now. 
\begin{figure}[t]
  \begin{CodeOut}
\footnotesize{
  \begin{alltt}      
circ = QuantumCircuit(1,1)
circ.x(0)

tomo = ProcessTomography(
       circuit=circ,
       measurement_basis=PauliMeasurementBasis(),
       measurement_qubits=None,
       preparation_basis=PauliMeasurementBasis(),\textit{ <-}
       preparation_qubits=None,
       basis_indices=None,
       qubits=None)
  \end{alltt}
}
\end{CodeOut}
\Caption{\label{fig:8}Example of call error}
\end{figure}

\subsubsection{QASM error (QE)} 
This detector detects problems with \CodeIn{qasm\_simulator} as the backend or when building \textit{qasm} programs with the Qiskit programming language.

\subsubsection{Discarded orders (DO)}
This detector determines if a deprecated method is being called, and it comes into play when an old operation or variable type is discarded due to a version update.

\subsection{Bug Detection}
The bug patterns shown in Table~\ref{table:1} represent the general bugs in quantum programs. In addition to syntactic bugs, it also contains some faulty logic in some quantum-related operations.
Based on the program information extraction modules (\CodeIn{QP\_Attribute}
and \CodeIn{QP\_Operation}),
as well as the detectors, we can perform a thorough static analysis for the quantum program files. To enable comprehensive and useful bug detection and benefit the debug process, we report the details of the buggy programs (code lines, content, etc.), bug types (patterns), and specific descriptions. We hope such information may help users improve the quality of quantum programs.

\section{Evaluation}
% 缺少检测时间的实验；以及precision，recall和F1的计算
%add environment:python3.10...
\label{sec:experiments}
This section presents the empirical performance of QChecker. We evaluate QChecker on Bugs4Q~\cite{zhao2021bugs4q}, which contains 42 real-world buggy quantum programs in Qiskit. The experiments were conducted on a server with an Intel i9-10940X CPU, 128G RAM, running on Ubuntu 20.04 with Python 3.10 installed.

% All the experiments are conducted on a server with CPU (Intel(R) Core(TM) i9-10940X CPU @ 3.30GHz), 128GB RAM, and Python version of 3.10. 
%Interestingly, we found that QChecker is very slight and requires minimal computation resources. Moreover, QChecker supports the main-stream operating systems such as MS-Windows, Linux, and MacOS.

\subsection{Metrics}
We adopt \emph{Precision}, \emph{Recall}, and \emph{F1-score} to evaluate the performance of QChecker. Specifically, for each bug \emph{b} in Bugs4Q, we use the bug type in Bugs4Q as ground truth and apply QChecker to its source quantum program. If the detection result of QChecker matches the corresponding bug type in Bugs4Q, we call this \emph{b} as \emph{True Positive} (TP). Otherwise, this \emph{b} is a \emph{False Positive} (FP). \emph{False Negative} (FN) is a ground-truth bug that can not be detected by QChecker. The \emph{Precision} is calculated as 
\begin{math}TP/ (TP + FP)\end{math}, \emph{Recall} as \begin{math}TP/(TP + FN)\end{math} and \emph{F1-score} as \begin{math}
2 \times Precision \times Recall / (Precision + Recall)
\end{math}. 
\subsection{Performance on Real-World Qiskit Programs}
\begin{figure}[h]
    \centering
    \includegraphics[width=0.5\textwidth]{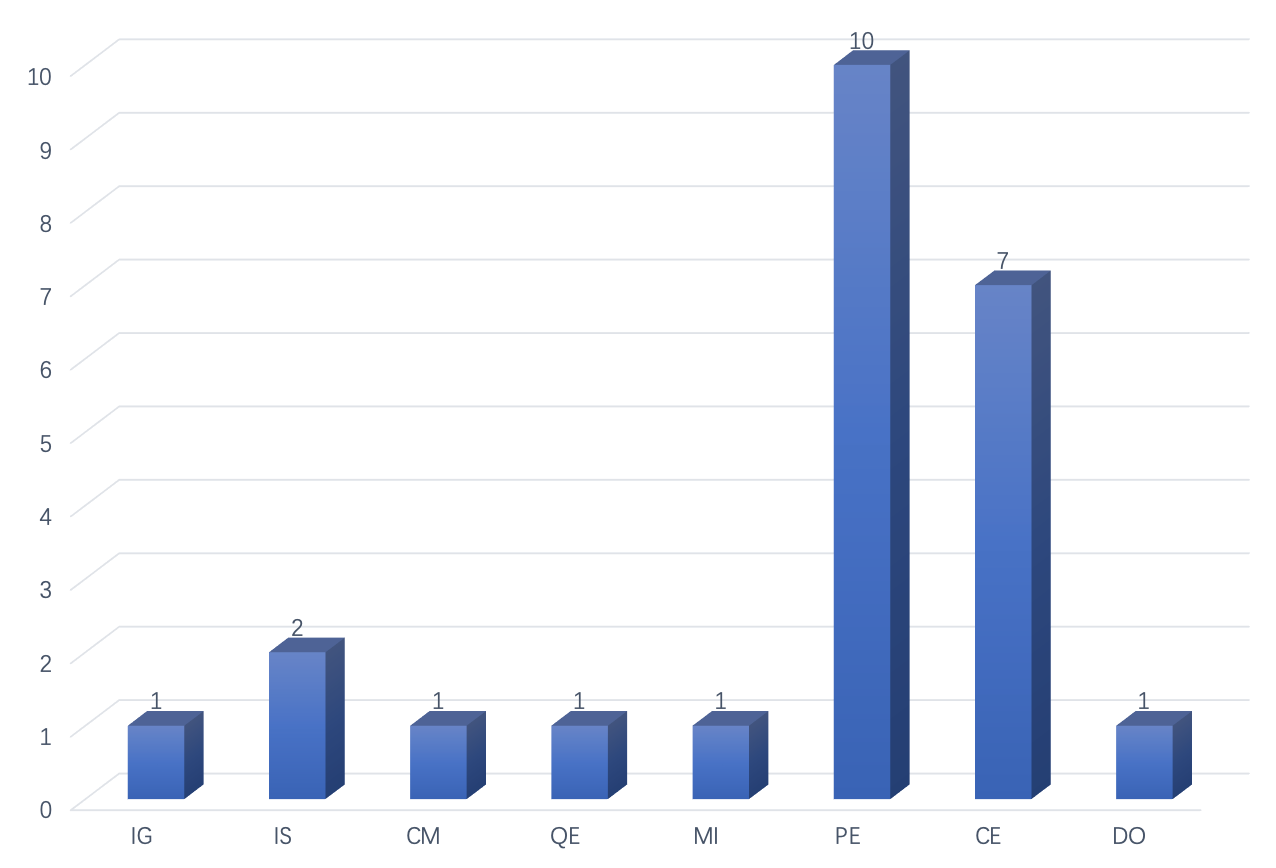}
    \caption{Distribution of bugs found by each detector.}
    \label{fig:9}

\end{figure}

First of all, there are 42 Qiskit programs detected by QChecker, and 24 bugs were found. 
Figure~\ref{fig:9} shows the empirical results produced by applying QChecker on Bugs4Q. 
Detector \CodeIn{PE} found 10 bugs, while \CodeIn{CE} and \CodeIn{IS} found 7 and 2 bugs, respectively. And other detectors found one bug in each. As we know, out of the 42 bugs in Bugs4Q, 22 bugs are output errors, i.e., the output of the program does not match the user's expectations. This condition makes Qchercker unable to detect these bugs. So we consider that the remaining number of bugs in Bugs4Q found by Qchecker is in line with expectations.
Combining Table~\ref{table:1} with the results in Figure~\ref{fig:9}, we analyzed and derived two points about the performance of each detector:
\begin{itemize}[leftmargin=2em]
\setlength{\itemsep}{3pt}

   \item \textit{The more cases the detectors can cover, the more bugs can be found}. As we made the detectors according to the relevance of the bug patterns, the number of cases covered by each detector may vary. Detector \CodeIn{PE} and \CodeIn{CE} can detect more cases than other detectors. And in fact, the two detectors made a better performance. 

   \item \textit{The performance of detectors also depends on the type of errors made by the programmers}. 
   Considering the limited number of bugs in Bugs4Q, we believe that some detectors perform poorly because they cover cases that rarely occur. For example, detector \CodeIn{IG}, \CodeIn{CM}, and \CodeIn{IS} can detect more than one case while the results are barely satisfactory. 
   %Therefore, we believe that most of the bugs come from parameter errors and call errors.
 \end{itemize}

 In addition, a detector that solves only one case does not indicate poor performance.
 Instead, with their increased functionality, these specialized detectors will realize their potential to detect bugs better.

\begin{table}
\centering
\caption{Performance of QChecker on Qiskit programs.}

\label{tbl:qc_performance}
\begin{tabular}{l|cccc}
\toprule
Performance & \emph{Prec.} & \emph{Recall} & \emph{F1-score} & \emph{Avg. Time}\\
\midrule
QChecker & 0.625 & 0.882 & 0.731 & 48.2ms \\
\bottomrule
\end{tabular}
\end{table}

The efficiency of QChecker on 42 Qiskit programs provided by Bugs4Q is shown in Table~\ref{tbl:qc_performance}. 
% Compared to static analysis tools for classical programs~\cite{li2019vbsac}, QChecker has achieved reasonable performance in detecting bugs in quantum programs. 
As illustrated in \cite{matteo2022morphq}, executing a quantum program on simulators can easily consume more than \begin{math}10^3\end{math} ms.
As a result, QChecker demonstrates high efficiency by taking an average time of only 48.2 ms to complete detection on a single quantum program.
%QChecker only takes 48.2 ms for complete detection on a single quantum program, which exhibits efficiency.
To better represent this, we investigated the 42 quantum programs in Bugs4Q.
The average execution time for these programs was 2.14 seconds, while the average amount of code per program was 31 LOC.
From Section~\ref{sec:background}, we already know that obtaining the state of a qubit requires a large number of iterations of the output to obtain its probability distribution, which we consider to be the main cost of executing one program.

In summary, QChecker has the advantage of being efficient and relatively effective in execution, while the disadvantage is QChecker does not avoid the problem of false positives.
As the number of qubits in future quantum programs increases, we believe it is necessary and effective to find bugs before program execution by means of static analysis.

\subsection{Extendability of QChecker}
%Although QChecker was developed for Qiskit programs, we tried to apply QChecker to other Python-based quantum programming languages.
%{\color{red}{Although QChecker is targeted at quantum programs written in Qiskit, we try to apply QChecker to seven buggy Cirq programs to evaluate its performance on other Python-based quantum languages. The results show that only one out of the seven bugs is detected by QChecker.}}
%Bugs4Q contains seven buggy programs written in Cirq. 
We next discuss the extendability of QChecker.
The example of a Cirq program and the corresponding detection result of QChecker are shown in Figures~\ref{fig:10} and \ref{fig:11}, respectively. After careful inspection of the results, we find that the information extraction part of QChecker can still function on other Python-based quantum languages (e.g., Cirq). The syntax difference between these quantum languages may cause the detectors designed for Qiskit fails to work, which leads to the lack of guaranteed performance.
%We can see almost information is extracted by QChecker so that the detectors can function.
%Between the seven buggy programs provided in Bugs4Q, one bug was found by the detector \CodeIn{IG}.
% From the evaluation, we consider QChecker can be used for common Python-based quantum programming languages.

In summary, the experimental results show that QChecker can successfully detect bugs in real-world Qiskit quantum programs, exhibiting the effectiveness of applying static analysis to quantum programs. %Although QChecker suffers significant performance degradation on Cirq programs, 
Besides, the intermediate results indicate that the QChecker can correctly extract the \CodeIn{QP\_Attribute} and \CodeIn{QP\_Operation} information from the underlying Cirq program, indicating that the QChecker can be easily extended to common Python-based quantum programming languages.

\begin{figure}[t]
\begin{CodeOut}
    \footnotesize{
  \begin{alltt}
    \textbf{import} cirq
    qubit = cirq.\textbf{NamedQubit}("myqubit")
    circuit = cirq.\textbf{Circuit}(cirq.H(qubit))
    \textbf{for} i \textbf{in} range(10):
        result2 = cirq.\textbf{measure}(qubit, key='myqubit')
        \textbf{print}(result2)
    \textbf{print}(circuit)
    result = cirq.\textbf{Simulator()}.simulate(circuit)
    \textbf{print}(result2)
  \end{alltt}
}
\end{CodeOut}
\Caption{\label{fig:10}An exapmle of Cirq program.}
\end{figure}

\begin{figure}[t]
\begin{CodeOut}
    \footnotesize{
    \begin{alltt}
    ('qubit', 'cirq.\textbf{NamedQubit}("myqubit")')
    ('circuit', 'cirq.\textbf{Circuit}(cirq.H(qubit))') 
    ('result', 'cirq.\textbf{Simulator}().simulate(circuit)')
    ('result2', 'cirq.\textbf{measure}(qubit,key="myqubit")')
    ==========================================
    'cirq.\textbf{NamedQubit}("myqubit")'
    'cirq.\textbf{Circuit}(cirq.H(qubit))'
    'range(10)'
    '\textbf{print}(circuit)'
    'cirq.\textbf{Simulator()}.simulate(circuit)'
    '\textbf{print}("result:")'
    '\textbf{print}(result2)'
    'cirq.\textbf{H}(qubit)'
    'cirq.\textbf{measure}(qubit,key="myqubit")',
    '\textbf{print}(result2)',
    'cirq.\textbf{Simulator()}'
\end{alltt}
}
\end{CodeOut}
\Caption{\label{fig:11}A Cirq program detected by QChecker}
\end{figure}

\section{Treats to Validity}\label{sec:threats}

\subsection{External Threats}
Even for the most widely used Qiskit quantum programming language, there are still not enough buggy programs. Moreover, existing quantum programs are usually run on simulators rather than on actual quantum computers, which leads to the small size of current quantum programs.
As a result, the number of bug patterns can be a threat to validity. We have put much effort into collecting bugs from quantum programs and extracting as many bug patterns as possible from these collected bugs. However, due to the limitation of the current scale of development and application of quantum programs, we cannot include more bug patterns in QChecker. 
%Nevertheless, the evaluation of Bugs4Q shows that QChecker can still detect bugs in various quantum programs.
Therefore, we will continue to collect quantum programs and their bugs, enrich QChecker's detection capabilities, and continuously update the tool. 

% The number of bug patterns can be a threat of validity.
% We have invested efforts to limited by the number of bugs in the quantum program, we .
% In addition, QChecker has not been evaluated extremely well. 
% Limited by the size of the quantum program, it may not take much time and resources to execute the quantum program, which does not reflect the value of QChecker.
% When large-scale quantum programs are available, QChecker can be well-evaluated.
\subsection{Internal Threats}
%QChecker supports detecting Python-based quantum programming languages (e.g., Qiskit, Cirq, ProjectQ). 
QChecker is designed for Qiskit and can be extended to other Python-based quantum languages (e.g., Cirq) with slight modifications.
However, it also has limitations. For instance, ProjectQ has overloaded the \begin{math}|\end{math} operator, which will cause the information extractor fails to work. As shown in Figures~\ref{fig:12} and \ref{fig:13}, QChecker could not extract information from the underlying ProjectQ program, such as the \CodeIn{H} gate and \CodeIn{CX} gate. These limitations will be resolved with the extension of QChecker.
%which leads to the detectors failing to work. 
%In addition, some of the detectors in QChecker could only cover a limited number of cases. 
%And the detectors can only provide error codes while not locating specific error statements.
%Finally, QChecker does not support detecting quantum programming languages which are not Python-based, such as Q\#. We leave these issues as future works for QChecker.
%In future work, we plan to integrate QChecker into Microsoft Visual Studio Code~\cite{code2019visual} as an extension and provide the corresponding features.

\begin{figure}[t]
\begin{CodeOut}
    \footnotesize{
  \begin{alltt}
    eng = \textbf{MainEngine}()
    qubits = eng.allocate_qureg(3)
    \textbf{H} | qubits[0]
    \textbf{CX} | (qubits[0], qubits[2])
    eng.flush()
    amplitudes = np.array(eng.backend.cheat()[1])
    amplitudes = np.abs(amplitudes)
    \textbf{All(Measure)} | qubits
  \end{alltt}
}
\end{CodeOut}
\Caption{\label{fig:12}An exapmle of ProjectQ program.}
\end{figure}

\begin{figure}[t]
\begin{CodeOut}
    \footnotesize{
  \begin{alltt}
    ('eng', '\textbf{MainEngine}()')
    ('qubits', 'eng.allocate_qureg(3)')
    ('amplitudes', 'np.array([1])')
    ('amplitudes', 'np.abs(amplitudes)')
    ==========================================
    '\textbf{MainEngine}()'
    'eng.allocate_qureg(3)'
    'eng.flush()'
    'np.array([1])'
    'np.abs(amplitudes)'
    '\textbf{All(Measure)}'
    'eng.backend.cheat()']
  \end{alltt}
}
\end{CodeOut}
\Caption{\label{fig:13}A ProjectQ program detected by QChecker.}
\end{figure}

\subsection{Verifiability}
This threat concerns the possibility of replicating this research.
%Considering the verifiability of this work,
we provide all the necessary details to help researchers replicate this work. The replication package is made publicly available at \url{https://github.com/Z-928/QChecker}.

\section{Related work}\label{sec:related}
% This section briefly discusses the following lines of closely related research.

A large number of error detection techniques based on static analysis~\cite{flanagan2002extended,hovemeyer2004finding,rutar2004comparison,shen2008xfindbugs,wang2022icpc} have been developed in classical software development. However, static analysis techniques for classical programs are difficult to apply directly to quantum programs due to the essential differences between quantum programs and classical programs. 
%In the following, we briefly discuss the static analysis and verification techniques for quantum programs.

Currently, static analysis techniques for quantum programs have emerged. Yu and Palsberg~\cite{yu2021quantum} proposed an abstract interpretation technique for quantum programs and used this technique to detect assertions to find errors in the programs. Xia and Zhao~\cite{xia2023static} proposed a practical static entanglement analysis technique to accurately analyze the entanglement information within and between modules in Q\# quantum programs, which can help find errors related to entanglement in the programs. 
ScaffCC~\cite{javadiabhari2015scaffcc} is a scalable compiler framework for the quantum programming language Scaffold~\cite{abhari2012scaffold}, which also supports entanglement analysis. ScaffCC explores data-flow analysis techniques to automatically track the entanglements within the code by annotating the output of the QASM-HL program, an intermediate representation of ScaffCC, to denote possibly entangled qubits. The analysis is conservative because it assumes that if two qubits interact, they are likely to have become entangled with each other. In contrast to these analysis methods, QChecker aims to find bugs in quantum programs based on bug patterns through static analysis.

Researchers have also extended Hoare logic to the quantum domain to support the formal verification of quantum programs~\cite{brunet2004dynamic,d2006quantum,ying2012floyd,unruh2019quantum,zhou2019applied}. Among them, Li {\it et al.}~\cite{zhou2019applied} introduced applied quantum Hoare logic (aQHL), which is a simplified version of quantum Hoare logic (QHL)~\cite{ying2012floyd}, with particular emphasis on supporting debugging and testing of the quantum programs. aQHL simplified QHL through binding QHL to a particular class of pre- and postconditions (assertions), that is, projections, to reduce the complexity of quantum program verification and provide a convenient way used for debugging and testing quantum software. However, as we know, formal verification can be costly and difficult to scale up.
%while our QChecker can be scaled up more easily.

Another line of research is to develop debugging~\cite{huang2019statistical,li2020projection} and testing~\cite{ali2021assessing,fortunato2022mutation2,honarvar2020property,long2022testing,wang2021generating} techniques for quantum programs, which are based on dynamic program analysis in principle. Although more accurate results can be obtained, the running cost of these techniques is relatively high compared to static analysis techniques due to the nature of quantum programs.

\section{Conclusion Remarks}\label{sec:conclusion}

%Existing static analysis tools fail to detect the potential bugs of quantum programs since quantum programs have significantly different properties from classical programs. 

This paper has presented QChecker, a static analysis tool for quantum programs to enable effective and efficient potential bug detection of quantum programs. QChecker involves two AST-based program information extraction modules and comprehensive bug detectors which can detect various bug patterns. We applied QChecker to the Bugs4Q benchmark suite and evaluated its effectiveness. The results show that QChecker can detect various types of bugs in quantum programs. 
In the future, we plan to extend QChecker to detect more bug patterns of Qiskit programs and support bug detection of other common quantum programming languages such as Cirq and ProjectQ. 
%Please refer to the code base\footnote{\url{https://github.com/Z-928/QChecker}} for implementation details.

%In the future, we plan to extend QChecker for more functionality, e.g., automatic test case generation, impact analysis, and metrics analysis. We will also incorporate more quantum simulation modules with bug detection. With the increasing number of quantum programming bugs, new bug types will be involved in QChecker, simultaneously. Moreover, QChecker will also be extended to detect program bugs in other common quantum programming languages.

% \section*{Acknowledgement} % not allowed in review
\section{Acknowledgment}
This work was supported by JST SPRING (Grant Numbers JPMJSP2136 and JPMJFS2132).

\bibliographystyle{IEEEtranS}
\bibliography{IEEEabrv,ref}

\end{document}

%% file: macro.tex
\usepackage{times}
\usepackage{graphicx}
\usepackage{epsf}
\usepackage{verbatim}
\usepackage{url}
\usepackage{color}
\usepackage{alltt}

\newcommand{\Caption}{\caption}

\newcommand{\CodeIn}[1]{{\small\texttt{#1}}}

\newcommand{\Comment}[1]{}

\newenvironment{CodeOut}{\begin{scriptsize}}{\end{scriptsize}}

\newcommand{\SmallSpace}{\vspace*{-1.4ex}}

\usepackage{xcolor}